\documentclass[sigconf,screen,nonacm,review=false,timestamp=false]{acmart}

\usepackage[mode=image|tex]{standalone}
\usepackage{tikz}

\usepackage[T1]{fontenc} 
\usepackage[utf8]{inputenc} 
\usepackage{pgfplots} 
\usepackage{amsmath}

\usepackage{xspace}
\usepackage[nolist]{acronym}

\usepackage{caption}
\usepackage{subcaption}

\usepackage[linesnumbered]{algorithm2e}
\usepackage{pifont}

\usetikzlibrary{arrows.meta}
\usepgfplotslibrary{statistics}
\usepgfplotslibrary{fillbetween}

\AtBeginDocument{%
  }

\setcopyright{none}

\acmConference[ASIACCS'24]{ASIACCS}{July 01--05, 2024}{Singapore}

\ifdefined\algorithmautorefname
\renewcommand{\algorithmautorefname}{Algorithm}
\fi 
\ifdefined\definitionautorefname
\renewcommand{\definitionautorefname}{Definition}
\fi

\begin{acronym}
\acro{CPU}{Central Processing Unit}
\acro{LRU}{Least-Recently Used}
\acro{RRP}{Random Replacement Policy}
\acro{PLRU}{Pseudo-LRU}
\acro{FRPLRU}{Random-Pseudo-LRU-Fixed}
\acro{DRPLRU}{Random-Pseudo-LRU-Dynamic}
\acro{VARP}{Variable Age Replacement Policy}
\acro{TLB}{Translation Lookaside Buffer}
\acro{RRIP}{Re-Reference Interval Prediction}
\acro{LLC}{Last-Level Cache}
\acro{FPGA}{Field Programmable Gate Array}
\end{acronym}
\acused{CPU}
\newcommand{\lru}{\ac{LRU}\xspace}
\newcommand{\rrp}{\ac{RRP}\xspace}
\newcommand{\frplru}{\ac{FRPLRU}\xspace}
\newcommand{\drplru}{\ac{DRPLRU}\xspace}
\newcommand{\varp}{\ac{VARP}\xspace}
\newcommand{\etal}{\textit{et al.}\xspace}
\newcommand{\ppp}{\textsc{Prime+\allowbreak{}Prune+\allowbreak{}Probe}\xspace}
\newcommand{\pp}{\textsc{Prime+Probe}\xspace}
\newcommand{\fr}{\textsc{Flush+Relaod}\xspace}

\begin{document}

\fancyhead[RO,LE]{\footnotesize ArXiv Version, 2023, December}

\title{On The Effect of Replacement Policies on The Security of Randomized Cache Architectures}

\author{Moritz Peters}
\orcid{0009-0008-8048-4020}
\affiliation{%
  \institution{Ruhr-University Bochum}
  \city{Bochum}
  \state{NRW}
  \country{Germany}
}

\author{Nicolas Gaudin}
\orcid{0000-0002-8044-3009}
\affiliation{%
  \institution{UMR 6285, Lab-STICC, Univ. Bretagne-Sud}
  \city{Lorient}
  \country{France}
}

\author{Jan Philipp Thoma}
\orcid{0000-0003-1613-732X}
\affiliation{%
  \institution{Ruhr-University Bochum}
  \city{Bochum}
  \state{NRW}
  \country{Germany}
}

\author{Vianney Lapôtre}
\orcid{0000-0002-8091-0703}
\affiliation{%
  \institution{UMR 6285, Lab-STICC, Univ. Bretagne-Sud}
  \city{Lorient}
  \country{France}
}

\author{Pascal Cotret}
\orcid{0000-0001-6325-0777}
\affiliation{%
  \institution{UMR 6285, Lab-STICC, ENSTA Bretagne}
  \city{Brest}
  \country{France}
}

\author{Guy Gogniat}
\orcid{0000-0002-9528-5277}
\affiliation{%
  \institution{UMR 6285, Lab-STICC, Univ. Bretagne-Sud}
  \city{Lorient}
  \country{France}
}

\author{Tim Güneysu}
\orcid{0000-0002-3293-4989}
\affiliation{%
  \institution{Ruhr-University Bochum}
  \city{Bochum}
  \state{NRW}
  \country{Germany}
}

\renewcommand{\shortauthors}{M. Peters, N. Gaudin, J.P. Thoma, V. Lapôtre, P. Cotret, G. Gogniat, T. Güneysu}

\begin{abstract}
Randomizing the mapping of addresses to cache entries has proven to be an effective technique for hardening caches against contention-based attacks like \pp.
While attacks and defenses are still evolving, it is clear that randomized caches significantly increase the security against such attacks.
However, one aspect that is missing from most analyses of randomized cache architectures is the choice of the replacement policy.
Often, only the random- and LRU replacement policies are investigated. 
However, LRU is not applicable to randomized caches due to its immense hardware overhead, while the random replacement policy is not ideal from a performance \textit{and} security perspective.

In this paper, we explore replacement policies for randomized caches.
We develop two new replacement policies and evaluate a total of five replacement policies regarding their security against \ppp attackers.
Moreover, we analyze the effect of the replacement policy on the system's performance and quantify the introduced hardware overhead.
We implement randomized caches with configurable replacement policies in software and hardware using a custom cache simulator, gem5, and the CV32E40P RISC-V core. 
Among others, we show that the construction of eviction sets with our new policy, VARP-64, requires over 25-times more cache accesses than with the random replacement policy while also enhancing overall performance.
\end{abstract}




\received{August 21, 2023}
\received[revised]{\today}
\received[accepted]{November 15, 2023}

\maketitle

\section{Introduction}
From a performance perspective, caches are a crucial component of modern \acp{CPU} that hide the high latency of main memory.
By storing frequently accessed data in fast SRAM memory close to the \ac{CPU} core, the processor can operate with fewer wait cycles for memory accesses.
Since the objective of caches is a reduction of the perceived memory access latency, low latency is a key characteristic of cache memory.
The \ac{CPU} must therefore be able to quickly determine whether the data associated with a given memory address is cached.
For this, most modern cache architectures are implemented as a set-associative structure, which uses part of the memory address to select a set of possible storage locations.
Set associative caches have established themselves as simple and performance-efficient solutions and are implemented in virtually all modern processors throughout the performance spectrum.

The timing difference between accessing cached data and uncached data is sufficiently large that it can be measured from user-level code.
Attackers have exploited this fact in various timing side-channel attacks to leak secret information from co-located processes.
For example, cache attacks have been used to leak secret keys from co-located processes~\cite{Osvik2006,Bernstein2005,CabreraAldaya2019,Yarom2014,Ronen2019,Mushtaq2020},
violate the isolation of secure enclaves~\cite{Brasser2017,Chen2019,Goetzfried2017}, and even as a key logger across co-located VMs~\cite{Gruss2016,Yarom2014}. 
Moreover, cache side channels are often used in the context of transient execution attacks like Spectre~\cite{Kocher2019} and Meltdown~\cite{Lipp2018}.
Generally, cache attacks can be grouped into flush-~\cite{Yarom2014,Gruss2016} and contention-based~\cite{Tromer2010,Osvik2006,DBLP:conf/uss/GrussSM15} attacks.
The former leverage a specific cache maintenance instruction, e.g., \texttt{clflush} on x86 that allows flushing an attacker-accessible address from the cache.
These attacks can easily be mitigated by not having an unprivileged cache flush instruction or duplicating read-only shared memory regions in the cache~\cite{Werner2019}.
On the other hand, contention-based attacks are much harder to mitigate since attackers can directly exploit the set-associative structure of caches. 
Proposed countermeasures include various partitioning-~\cite{Zhou2016,Liu2016,Kiriansky2018,Sanchez2012,Kim2012} and detection~\cite{Chen2014a,Chen2014,Fang2018,Yan2016} schemes. 
However, a particularly promising approach preventing contention-based cache attacks is index-randomization~\cite{Canale2023,DBLP:conf/micro/Qureshi18,DBLP:conf/isca/Qureshi19,DBLP:conf/uss/SaileshwarQ21,DBLP:conf/sp/SongLXLWL21,DBLP:conf/ndss/TanZB020,Thoma2023,DBLP:conf/isca/WangL07,Werner2019}. 
In index-randomized caches, the address-to-index mapping is no longer determined by a fixed physical address range.
Instead, a randomization function~\cite{Canale2023} selects pseudorandom entries in each cache way for a given address.
This renders contention-based cache attacks immensely complex for caches protected purely by index randomization~\cite{Werner2019,purnal2021systematic} and infeasible for more sophisticatedly hardened cache architectures~\cite{DBLP:conf/uss/SaileshwarQ21,Thoma2023}.

A major oversight in current randomized cache architectures is the choice of the replacement policy which we aim to address in this work.
Many works propose choosing between \ac{LRU}- and random replacement policy for the randomized cache design.
However, we find that in practice, the \ac{LRU} policy is unsuitable for randomized caches as the hardware would need to be able to compare the age of any entry to any other entry, resulting in an extremely high overhead in hardware. 
Moreover, we identified that \ac{PLRU} policies are not simply transferable to randomized cache architectures.
The random replacement policy is not favored on traditional \acp{CPU} due to reduced performance compared to more sophisticated policies.
Moreover, the effects of the replacement policy on the overall security of randomized caches and their resistance against \ppp -style attacks~\cite{purnal2021systematic} have only been dealt with at the surface so far.
In this work, we primarily investigate replacement policies for use in randomized cache architectures.
We propose two new replacement policies and investigate a total of five replacement policies in the context of randomized caches.
Besides the impact on system performance, we investigate the effect on the security against \ppp attackers and the overhead in hardware.
For this, we use software-based simulation as well as an \ac{FPGA}-based implementation on a RISC-V core.

To summarize, the contributions of the papers are:
\begin{itemize}
\item We analyze suitable replacement policies for randomized cache architectures and propose two new replacement policies.
\item We investigate the effects of the replacement policies on the security of the randomized cache architecture. Therefore we assume a \ppp~\cite{purnal2021systematic} attacker, which is the most recent generic attack against randomized caches. 
According to our findings we reveal that a random replacement policy is the worst choice and our new proposal VARP-64 increases the complexity of constructing eviction sets by a factor of at least 25.
\item We implement a randomized cache with the considered replacement policies in a high-level cache simulator, the software-based \ac{CPU} simulator gem5~\cite{lowepower2020gem5}, and in hardware on a CV32E40P RISC-V core. We leverage these evaluation platforms to analyze the security, performance, as well as hardware area overhead of the considered replacement policies.
We find that all investigated replacement policies perform better than \rrp and secure replacement policies can be implemented with hardware costs similar to \rrp.
\end{itemize}

\section{Background}
In this section, we introduce background information on caches and establish the terminology used in this paper.
Further, we introduce common cache attacks and cache randomization as one possible countermeasure. 

\subsection{Terminology}
We first need to establish a terminology that is used throughout the paper.
It can be found in \autoref{rprc:tab:terminology}.
For the cache parameters, we denote the number of cache sets as $S$, the associativity as $W$, and the entry size as $B$.
Furthermore, we denote an eviction set as $G$, the initial pruning set used in \ppp as $k$, and a target or victim address as $V$.

\begin{table}
\footnotesize
    \centering
    \caption{Terminology} \label{rprc:tab:terminology}
    \begin{tabular}{ll | ll}
        \toprule
        Cache Sets & $S$  &
        Cache Associativity & $W$ \\
        Cache Entry Size & $B$ &
        Eviction Set & $G$ \\
        Initial Pruning Set & $k$ &
        Victim Address & $V$ \\
        \bottomrule
    \end{tabular}
\end{table}

\subsection{Caches}
Accessing data from the main memory is slow.
Therefore modern systems use caches to prevent the CPU from stalling on requested data.
They store data that is frequently accessed to speed up repetitive lookups.
Newer systems use entire cache hierarchies consisting of three cache levels.
These range from the L1 cache, which is the fastest but also the smallest, to the L3 cache, which is the largest but also the slowest.
The L3 cache is also called the \ac{LLC} and is usually shared across cores.
When the CPU requests data at a particular address, it first queries the caches using that address.
If the data is found in the cache, it gets sent directly to the CPU.
This is referred to as a cache hit.
Conversely, the request is passed to the main memory if the data is not found in the cache.
This is referred to as a cache miss.
The CPU must then wait for the main memory to respond with the requested data.
Requests always start at the lowest cache level (L1).
If a cache miss occurs, the request is moved up one level until it reaches the main memory.

Data in the cache is stored in the form of a cache line.
A cache line comprises $B$ bytes, usually $B=64$.
In order to find data for a specific memory address among the $B$ bytes in a cache line, the lowest $log_2(B)=6$ bits of the address are used as an offset.

Most recent CPUs implement \textit{set-associative} caches.
These are organized in a table-like structure with $W$ ways (columns) and $S$ sets (rows).
A cache line now maps to a set in the cache and can be placed in any of the $W$ ways for that set.
The memory address is divided into offset, set index, and tag bits to select the set.
The lower $log_2(B)$ bits are used as the offset, followed by the $log_2(S)$ set bits and then the tag bits.
As in the \textit{fully-associative} cache, a replacement policy chooses a cache line for eviction if all ways in a cache set are occupied.

\begin{figure}[t!]
    \centering
    \includegraphics[width=.9\columnwidth]{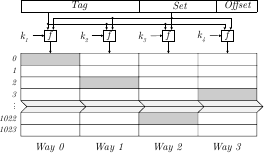}
    \caption{A randomized cache with 4 ways and 1024 sets.}
    \label{rprc:fig:randcache}
\end{figure}

\subsection{Attacks on Caches}
Many different cache attacks have been proposed over the past years.
The attacker aims to observe cache access patterns of co-located processes to extract secret data like cryptographic keys. 
He exploits the timing difference between cache hits and misses to achieve this.
The most prevalent attacks are \textsc{Flush+Reload}~\cite{Yarom2014}, \textsc{Evict+Reload}~\cite{DBLP:conf/uss/GrussSM15}, and \textsc{Prime+Probe}~\cite{DBLP:conf/sp/LiuYGHL15}.
\textsc{Flush+Reload} is a flush-based attack that requires shared memory between the attacker and the victim and an instruction to flush an address from the cache.
The target address is flushed from the cache hierarchy in the first step.
The attacker then triggers the victim and observes whether the target address is now loaded from the cache or main memory.
\textsc{Evict+Reload} works in the same way as \textsc{Flush+Reload}.
However, instead of the \texttt{clflush} instruction, an eviction set is used to evict the target line from the cache.
\textsc{Prime+Probe} does not rely on the \texttt{clflush} instruction nor requires shared memory between the attacker and the victim.
First, an eviction set is used to prime the cache and evict the target address.
Next, the attacker waits for the victim to make its access.
Finally, the attacker again accesses the eviction set and measures the access latencies to identify the victim's access.
Since \textsc{Evict+Reload} and \textsc{Prime+Probe} do not use a specific flush instruction but an eviction set, these fall into the category of contention-based cache attacks.

\subsection{Randomized Caches}
While flush-based attacks can be mitigated by making the flush instruction privileged or duplicating read-only shared addresses in the cache~\cite{Werner2019}, preventing contention-based attacks requires more sophisticated approaches.
One such approach is cache randomization~\cite{Canale2023,DBLP:conf/micro/Qureshi18,DBLP:conf/isca/Qureshi19,DBLP:conf/uss/SaileshwarQ21,DBLP:conf/sp/SongLXLWL21,DBLP:conf/ndss/TanZB020,Thoma2023,DBLP:conf/isca/WangL07,Werner2019}.
More specifically, the address-to-set mapping is randomized.
This makes it harder for an attacker to construct eviction sets for a given target address efficiently.
\autoref{rprc:fig:randcache} shows an example of a randomized set-associative cache with four ways and 1024 sets.
A pseudorandom function $f$ takes the memory address as an input and returns a pseudorandom cache set index for each cache way.
The gray entries indicate that they were pseudorandomly selected by $f$.
In this specific example, $f$ uses a key to generate different set indices in every way.
Note that the function $f$ needs to be called before every cache lookup.
Hence, the latency of $f$ must be as low as possible.
\section{\ppp}
\label{rprc:sec:ppp}

Purnal \etal \cite{purnal2021systematic} studied the security of randomized cache architectures against contention-based cache attacks.
Though randomization significantly increases the effort for such attacks, they found that performing targeted evictions of a victim address is still possible.
For this, they use a generalized eviction set $G$ which contains \textit{partially congruent} addresses, i.e., addresses that collide with the victim address in at least one cache way.

Prunal \etal demonstrated that it is possible to efficiently compute partial congruent eviction sets that have a high probability of causing collisions with a victim address $V$.
\autoref{rprc:alg:ppp} gives an overview of their proposed eviction set construction algorithm dubbed \ppp.
In the following, we mark occasions where the attacker needs to observe an eviction of their own addresses (i.e., \textit{catching an access to $V$}) with \ding{182} and occasions where the attacker needs to evict a targeted address with \ding{183}.
These will be relevant for the security analysis in \autoref{rpcp:sec:security}.

Similar to the traditional \pp algorithm, first, a large set of addresses $k$ is accessed to populate the cache.
Next, in the prune step, these addresses are reaccessed.
Both actions are covered by the loop starting in \autoref{rprc:alg:ppp:pp}.
Pruning is necessary since, for large $k$, addresses in $k$ may evict each other during the initial prime step.
If this step were skipped, the attacker could not distinguish an eviction caused by the victim from evictions caused by addresses in $k$.
Pruning is repeated until no more self-evictions occur.
Next, in \autoref{rpcp:alg:ppp:victim_access}, the victim is triggered to access $V$.
Then, in \autoref{rprc:alg:ppp:detect}, the attacker measures the latency of accessing each address in $k$ again (\ding{182}).
When a cache miss is detected on one of these addresses, the attacker learns that the address was evicted by $V$ and, thus, that it collides with $V$ in at least one cache way.
Finally, $V$ needs to be evicted from the cache to start the next iteration of the eviction set assembly (\ding{183}).
While $V$ may be evicted by accessing the next $k$, Purnal \etal propose to access the addresses in $G$ again before starting the next iteration (see \autoref{rprc:alg:ppp:evict}).
This increases the probability of evicting $V$ since the addresses in $G$ are known to collide with $V$.
This process is repeated until enough addresses are added to $G$.

\begin{algorithm}[ht]
\small
    \DontPrintSemicolon
    \KwIn{Victim $V$, Eviction Set Size $X$, Pruning Set $k$}
    \KwOut{Eviction Set $G$}
    \SetKwComment{Comment}{// }{}
    \SetKwProg{Fn}{Function}{:}{end}
    \SetKwFunction{Fck}{generate_ev}
    \SetKwFunction{Faccess}{access}
    \SetKwFunction{Fpp}{prime+prune}
    \SetKwRepeat{Do}{do}{while}

    \While{$|G| < X$}{
        \Comment{Prime and Prune}
        \Do{one access misses the cache}{\label{rprc:alg:ppp:pp}
            \ForEach{addr in k}{
                \Faccess{addr}\;
            }
        }
        \BlankLine
        \Comment{Trigger victim access}
        \Faccess{V}\;\label{rpcp:alg:ppp:victim_access}
        \BlankLine
        \Comment{Probe (\ding{182})}
        \ForEach{addr in k}{\label{rprc:alg:ppp:detect}
            is\_miss = \Faccess{addr}\;
            \If{is\_miss}{
                $G = G \cup \{addr\}$
            }
        }
        \BlankLine
        \Comment{Try to evict $V$ (\ding{183})}
        \ForEach{addr in G}{\label{rprc:alg:ppp:evict}
            \Faccess{addr}\;
        }
    }
    \caption{Algorithmic overview of the \ppp attack~\cite{purnal2021systematic}.}
    \label{rprc:alg:ppp}
\end{algorithm}

Once a sufficiently sized $G$ is constructed, the attack can be carried out similarly to \pp: The attacker first accesses $G$ to bring the addresses into the cache. 
Then, the victim is triggered, which causes a secret-dependent cache access to $V$.
The attacker learns if $V$ was accessed by probing the addresses of $G$.
One important difference to \pp is that $G$ cannot be used to evict $V$ from the cache after this. 
That is because all but one addresses of $G$ are already cached.
Since the cache is randomized and addresses are mapped to multiple locations in the cache, the probability that the evicted address of $G$ evicts $V$ during the probe step is low.
Therefore, the attacker must rely on different approaches to evict $V$ (\ding{183}).
One way would be to create another eviction set $G'$ for the sole purpose of evicting $V$.
Another option would be to access a large number of random addresses to evict $V$ probabilistically.
However, both approaches have their drawbacks.
Creating $G'$ increases the required profiling effort during the eviction set construction.
Using such a $G'$ later in an actual attack will be faster since fewer accesses are required to evict $V$.
Using random addresses reduces the profiling overhead during the eviction set construction.
However, the attacker is required to access more arbitrary addresses to evict $V$ during the attack than when using $G'$.
We analyze this trade-off for each replacement policy in \autoref{rpcp:sec:security}.

\section{Replacement Policies}\label{replacementpolicies}
Until now, replacement policies for randomized cache architectures have been insufficiently investigated.
Most randomized cache architectures provide a choice of replacement policy between \lru and random replacement~\cite{Thoma2023,purnal2021systematic,Werner2019}.
However, \lru is entirely impractical on randomized caches since the cache controller would need to be able to compare the age of every cache entry against every other.
For this, either the timestamp used for the \lru decision must be immensely large, or, on every access, the replacement data of every single cache entry must be updated.
For example, if we want a true \lru cache with full access ordering, we would need $\lceil{}log_2(W!)\rceil{}$ bits per way.
While this might be feasible for small associativities, using larger associativities make this impractical.
Additionally, this would be even worse in a randomized cache because we would need to keep the full access ordering for all entries in the cache.
Since the replacement policy is implemented in hardware and low latency and low memory overhead are key characteristics, this is not feasible.
Even traditional \ac{PLRU} approaches do not apply to randomized cache architectures since the set of compared entries is randomly selected, and the cache controller does not know and cannot store the age relation between a random selection of entries.

The randomized \ac{TLB} design TLB\-Coat~\cite{DBLP:journals/tches/StolzTSG23} introduces RPLRU, which attempts to approximate \lru. 
Since RPLRU was initially proposed in the context of \acp{TLB} and the security of the replacement policy itself was not the main focus of this work, a detailed security- and performance analysis in the context of caches is currently missing.
Moreover, RPLRU represents only one of several possible solutions for such a replacement policy.
In this paper, we want to gain a more detailed insight into the effect of design decisions and, therefore, analyze other possible solutions for cache replacement policies in randomized caches that approximate \lru.

In the following, we analyze the security of different replacement policies in the context of randomized caches.
Therefore, we investigate the replacement behavior in an attack scenario.
In this section, we first describe the replacement policies investigated in this paper.
In \autoref{rpcp:sec:security}, we analyze the security of each replacement policy against random and targeted eviction approaches.

\noindent\textit{\textbf{1. \acf{RRP}:}}
\rrp is a straightforward policy as it replaces cache lines at random.
When a cache line needs to be evicted, the (pseudorandom) indexing function provides a set of $W$ possible candidate entries for replacement.
The cache controller randomly chooses one of the candidate entries for replacement.
Though generally speaking, the generation of randomness in hardware is expensive, the amount of randomness required is low, and the limited observability of the attacker allows the use of pseudorandom numbers.
\rrp is the only stateless replacement policy considered in this paper.
That is, the cache controller is not required to store meta information for the replacement alongside the cache entry.
Due to its simplicity, \rrp is often mentioned in the context of randomized caches~\cite{Thoma2023,purnal2021systematic,Werner2019}.
In addition, the intuition is that increasing the randomization space would benefit the security of randomized caches.
In our security analysis (see \autoref{rpcp:sec:security}), we show that this intuition does not necessarily hold.

\noindent\textit{\textbf{2. \acf{LRU} Replacement Policy:}}
The \lru policy replaces the least recently used cache line from the candidate entries.
Approximations of \lru are often used in real-world caches since it maps well to the temporal locality of most workloads, i.e., addresses that have been accessed recently will likely be reaccessed in the short term.
Therefore, the cache controller needs to store information about the order of accesses for each entry. 
On a cache miss, the oldest entry of the set will be replaced.
The newly placed entry becomes the most recently used one in the target set, and the ages of all other entries in this set must be adjusted accordingly.
Similarly, on a cache hit, the accessed entry becomes the most recently used, and all other set entries are adjusted.
To enable efficient hardware implementations of \ac{LRU}, approximations like tree-\ac{PLRU} exist~\cite{Robinson2004}.
Even though the policy is not suited for randomized cache architectures outside of simulated environments, we include \ac{LRU} in our experiments to investigate how the security would be affected if a randomized cache \textit{could} use this replacement policy.

\noindent\textit{\textbf{3. \acf{DRPLRU} Replacement Policy:~}}
\begin{figure}[ht!]
    \centering
    \includegraphics[width=.8\columnwidth]{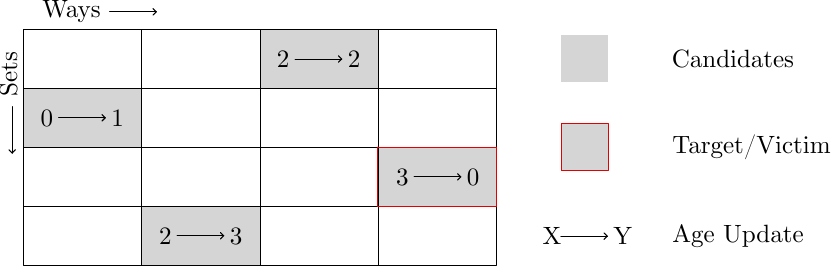}
    \caption{The \drplru replacement policy. The target is selected from the candidates based on an unsynchronized age indicator of width $W$. If entries have the same age, one of the oldest entries is selected randomly. The candidates are ordered based on their ages and updated accordingly.}
    \label{rprc:fig:drplru_cache_hitmiss}
\end{figure}

\noindent In an effort to approximate the \lru policy on randomized caches, we designed \drplru, a pseudo-\ac{LRU} policy.
Upon access, the cache selects a set of $W$ replacement candidates, each associated with an age indicator between 0 and $W-1$. 
\drplru then selects the entry with the highest age indicator from the candidates for replacement. 
Since the candidates are chosen as a pseudorandom subset with size $W$ of all entries, it is not guaranteed that the ages of the candidate entries are unique for comparison.
Hence, if two or more entries share the same age, \drplru selects a random entry with maximum age; i.e., when two entries have age $W-1$, \drplru selects one randomly.
Finally, the age of the candidate entries is updated to be ordered from 0 to $W-1$.
The new entry thereby is assigned with age $0$.
Again, if two entries have the same age, their respective order is selected randomly.
The process is visualized in \autoref{rprc:fig:drplru_cache_hitmiss}.
Like \lru, the accessed element becomes the most recently used on a cache hit, and the other entries in the same dynamic set are ordered.

The intensive behind \drplru is to preferably replace older entries.
For recently used entries, their age is low, and thus, they are less likely to be replaced by a new entry.
By forcing an order to the candidate entries after each access, we ensure that the ages do not converge towards the maximum over time, essentially resulting in an \rrp policy.

\noindent\textit{\textbf{4. \acf{FRPLRU} Replacement Policy:~}}
\begin{figure}[ht!]
    \centering
    \includegraphics[width=.8\columnwidth]{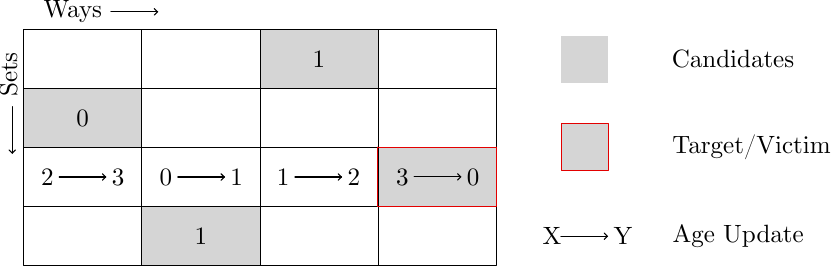}
    \caption{The \frplru replacement policy. The target is selected from the candidates based on the age of the candidates relative to other entries that share the set index. The age is only updated in the cache set of the replaced entry.}
    \label{rprc:fig:frplru_cache_hitmiss}
\end{figure}

\noindent The randomized \ac{TLB} design TLBCoat~\cite{DBLP:journals/tches/StolzTSG23} proposed using a different approximation of \lru for randomized caches dubbed RPLRU.
In this paper, we will refer to it as \frplru.
Similar to \drplru, \frplru assigns an age indicator to each cache entry.
These age indicators represent the age of a cache line relative to other entries stored at the same index, i.e., the cache set in the non-randomized setting.
Hence, in a $W$-way cache, the age indicators range from 0 to $W-1$. 
When a cache line needs to be evicted, the indexing function provides $W$ pseudorandom cache lines as eviction candidates.
\frplru selects the cache line with the highest age to be evicted from these.
If multiple cache lines share the same age, one is chosen randomly.
Thus, similar to \rrp, the cache controller needs to be able to generate pseudorandom numbers.
Until now, \frplru has been similar to \drplru.
However, instead of adjusting the ages of the dynamically selected replacement candidates, \frplru adjusts the ages within the set of entries stored at the same cache index (i.e., the cache set in the non-randomized case).
The newly inserted entry is assigned the smallest age, and the ages of all cache entries sharing the index are adjusted accordingly, as visualized in \autoref{rprc:fig:frplru_cache_hitmiss}.
An access to a cached cache line resets its age to the smallest one possible, and all ages of the cache lines in the same set are again adjusted accordingly.
This implies that no two cache lines share the same age in one cache set.
An advantage of \frplru is that traditional \ac{PLRU} techniques can be used on the cache set to maintain the ages.

\noindent\textit{\textbf{5. \acf{VARP}:~}}
\begin{figure}[ht!]
        \centering
        \includegraphics[width=.8\columnwidth]{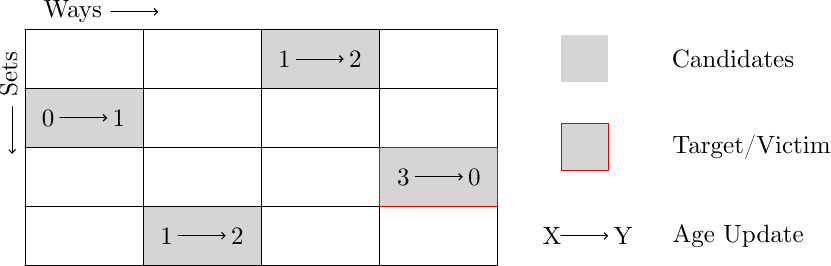}
        \caption{The \varp replacement policy. The target is selected from the candidates based on an unsynchronized age indicator of variable width. If entries have the same age, one of the oldest entries is selected randomly. The accessed entry is set to 1, and all other entries' ages are increased.}
        \label{rprc:fig:varp_cache_miss}
\end{figure}

\noindent \varp uses unsynchronized age indicators in each cache entry.
Hence, these age indicators are not ordered relative to cache lines in the same set or those selected by the indexing function.
In contrast to the other replacement policies, \varp can be configured to use a different number of ages.
Specifically, the ages for \varp range from 0 to $m-1$, where $m$ is the highest possible age (e.g., 64).
The idea behind this configurable age limit is to explore the effects when using, for example, only four possible ages in contrast to allowing cache lines to have up to 1024 different ages.
The results are discussed in \autoref{rpcp:sec:security} and detailed in \autoref{rprc:apx:varp_catch}.

On a cache miss, the indexing policy selects $W$ candidate entries.
\varp selects the entry with the highest age of the candidates. 
Since the ages are not organized in any way, there might be multiple candidate entries with the same age.
Like the other replacement policies, one of these entries is selected randomly.
The new entry's age is set to 0, and the age of all remaining candidate entries is shifted left by one.
On a cache hit, the accessed entries' age is reset to the lowest value while the ages of the other entries in the dynamic set are shifted left by one.
\autoref{rprc:fig:varp_cache_miss} visualizes the replacement policy.

\section{Evaluation Environment}
\label{rprc:sec:eval_env}
In this section, we introduce our evaluation environment.
First, the software cache simulator and its implementation are presented.
It is extensively used for security analysis with \ppp.
Second, we describe how we used the gem5 simulator \cite{lowepower2020gem5} to analyze the performance of the replacement policies discussed in this paper.
Last, we describe our hardware implementation, which we used to derive area results for SCARF \cite{Canale2023} and the replacement policies.

When choosing a cache randomization function, two main requirements must be satisfied.
First, the function needs low latency to avoid becoming the bottleneck for cache accesses.
Second, an attacker should not be able to bypass the randomization by, for example, being able to compute the output of the randomization function.
In our environment, we choose to use SCARF \cite{Canale2023} as a randomization function for our randomized cache.
It satisfies both requirements by being low-latency and cryptographically secure with respect to the cache attacker model.
SCARF is a 10-bit tweakable block cipher specifically designed for cache randomization.
It takes the 48-bit tag and 10-bit set index as tweak and plaintext, respectively, in combination with a 240-bit key to encrypt the set index.

The randomized cache is implemented as shown in the beginning of this paper in \autoref{rprc:fig:randcache}, where each way uses a separate key.
For simplicity, the keys are static and not changed during runtime.
In a real environment, one would choose to renew the keys periodically to increase security further.

We use the same cache configuration for all of our three environments.
The cache is set-associative with 1024 sets and four ways.
The index-to-set mapping is randomized using SCARF \cite{Canale2023}.
Each cache entry can hold 64 bytes.

\noindent\textit{\textbf{Software-Cache Simulator.}}
The software-based simulator is written in C++ and enables fast- and noiseless simulation of the cache state and replacement policy.
Compared to gem5~\cite{lowepower2020gem5}, it allows simple examination of the internal state and is much faster due to the focus on the cache only.
By translating memory accesses of our attack code to function calls to the simulator, we can realistically simulate and monitor the cache behavior of the attack.
The cache simulator offers various configuration opportunities, including the type of cache (randomized vs. non-randomized), the size, the associativity, and the replacement policy.
The replacement policies evaluated in this paper are implemented in separate modules.
Each module implements the same interface to enable the replacement policies to be used interchangeably.

The cache itself is implemented as a two-dimensional array of cache entries.
Each cache entry stores a valid bit flag, a tag, and a replacement state.
The interface of the cache simulator includes an \texttt{Access()} and a \texttt{Flush()} function.
A call to \texttt{Access()} with an address causes the cache to extract the cache set index and tag and provides the replacement policy with a list of entries suitable for the address to be cached.
The replacement policy will select one of the entries depending on the replacement data.
When a suitable entry is selected, its valid bit is set, the tag is stored, and the replacement policy updates the replacement data.
Since our cache simulator has no distinct time difference between a cache hit and a cache miss, the \texttt{Access()} function returns true if the address was cached.
Flushing an address from the cache with \texttt{Flush()} causes the cache to search for the corresponding entry and sets its valid bit to zero if it is found.

\noindent\textit{\textbf{Gem5 Implementation.}} 
We decided to use the gem5~\cite{lowepower2020gem5} for the performance analysis of the replacement policies on a large-scale processor.
It allows us to simulate a complete Linux system with a randomized cache that implements our replacement policies.
In addition, gem5 provides accurate statistics on cache hits and misses and IPC values, which are critical metrics for evaluating replacement policies.

We implemented a randomized cache architecture to gem5 using SCARF as randomization function.
Hence, the number of sets is restricted to 1024.
Gem5 comes with \lru and \rrp as cache replacement policy options by default.
We implemented the remaining replacement policies and made them configurable from the gem5 configuration file.

\noindent\textit{\textbf{Hardware Implementation.}}
We describe our different replacement policies using the SystemVerilog language in order to evaluate their area.
We test our implementations on our system, which is a mono-core CPU described in SystemVerilog.
The CPU embeds one RISC-V core and one level data cache.
We chose the CV32E40P core implementing the RISC-V Instruction Set Architecture.
CV32E40P core is a 4-stage in-order core that meets the requirements of embedded systems.
It physically accesses the memory using 22-bit addresses.
The cache is a 4-way set-associative with 1024 sets and 16-byte cache lines. 
The L1 data cache uses SCARF to randomize the cache access between sets for security purposes.
One SCARF module is dedicated per way.
As mentioned in Figure \ref{rprc:fig:randcache}, we implement one SCARF per way (4: in our setup).
Each SCARF uses a unique key that is randomly defined during implementation.
The selected set results from a key (240 bits), the tag, and the index of the accessed address.
We then choose a way depending on the replacement policy implemented in the cache respecting the aforementioned specifications.
Except for RRP, we need to store the states of each cache line.
For DRPLRU and FRPLRU, this corresponds to 2 bits per cache line, i.e., 8192 bits (1024 sets, 4 ways).
VARP requires 6 bits per cache line, giving a total of 24576 bits.
For the RRP and the randomness needed in \drplru, \frplru, and \varp, we use a 16-bit LFSR (Left Feedback Shift Register).
Due to the use of SCARF, certain cache lines selected by the SCARF modules may be part of the same set.
To remedy this, the replacement policies pick a way as follows: first, we check for the oldest cache line and count the number of ways with the same age.
Then, we select the way with the highest age.
If there are multiple, we randomly select the way among the highest age ways through a binary tree controlled by the LFSR.
The states are updated following the specifications mentioned in Section \ref{replacementpolicies}.

\section{Security}
\label{rpcp:sec:security}

In the following, we systematically evaluate how the presented replacement policies affect the security of randomized caches.
Therefore, we primarily focus on \ppp attacks, the state-of-the-art attack vector against randomized cache architectures.


\subsection{Attacker Model}
Consistent with the attacker model for randomized cache architectures~\cite{Thoma2023,purnal2021systematic,Werner2019}, we assume that the attacker and the victim simultaneously operate on the same CPU with a shared cache.
Moreover, we assume the victim executes code that operates on a secret value the attacker wants to leak.
Depending on the secret, the victim program may access an address exclusively mapped to the victim's address space.
Hence, we are in the \pp attacker model, where the attacker cannot access the target address.
If the target address is accessible to the attacker, \fr attacks are trivial, and randomization provides no measures to increase the security against them.
The attacker's goal is to construct a small eviction set that has a high probability of catching an access to the target address.
In non-randomized caches, eviction sets are equally helpful to \textit{a)} evict the victim address from the cache and \textit{b)} observe an access by the victim.
In fact, the same operation, i.e., priming the cache set, fulfills both purposes.
On randomized cache architectures, however, depending on the replacement policy, the probability of an eviction set \textit{replacing the victim} (\ding{183}) does not equal the probability of an eviction set address \textit{being replaced by the victim} (\ding{182}).
In most attack scenarios, the attacker first needs to evict the victim address from the cache and then requires the victim address to replace one of the eviction sets addresses.

We define the attacker's goal to force the eviction of a given target entry (\textit{``victim''}) by accessing other addresses. 
The naive approach for the attacker would be to fill the cache with random data until the entry is evicted. 
A more targeted attack would use a \ppp -style attack~\cite{purnal2021systematic}, using colliding addresses to evict the victim address.

\subsection{Catching Accesses (\ding{182})} \label{rprc:sec:catching_accesses}
The ability to observe (\textit{catch}) victim accesses to the cache is a basic requirement of \ppp. 
During the eviction set construction, the attacker needs to be able to catch accesses after priming (and pruning) the cache using random addresses.
In the attack phase, the attacker must catch an access after priming and pruning the generalized eviction set $G$.
Naturally, the probability of catching an access to $V$ depends on the number of primed addresses, i.e., $|k|$ during the eviction set construction and $|G|$ in the attack phase.
In this section, we analyze the effect that different replacement policies have on the catch probability.
Note that though we show the results for a specific cache configuration (i.e., 4 ways, 1024 sets), we verified that the characteristics shown in the following are similar for other reasonable cache configurations.
For the evaluation, we use the  cache simulator described in \autoref{rprc:sec:eval_env}.
First, we analyze the size of $G$ required to reliably catch an access to $V$ for our replacement policy candidates.

\begin{figure}[ht!]
    \centering
    \includegraphics[width=\columnwidth]{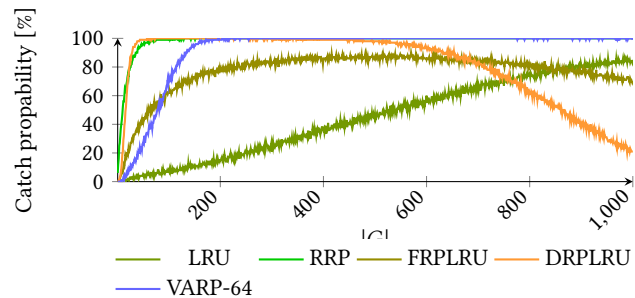}
    \caption{Catching probability for different replacement policies as a function of the eviction set size $|G|$.}
    \label{rprc:fig:catch_prop_g}
\end{figure}

\autoref{rprc:fig:catch_prop_g} shows the average catching probability of several eviction set sizes $|G|$ for each replacement policy.
For each $|G|$, we construct a generalized eviction set $G$ which is then primed and pruned.
This is essential to detect the access of $V$ rather than self-evictions within addresses of $G$.
Now $V$ is accessed.
If we reaccess $G$ and find a cache miss, the access to $V$ has been caught.
Finally, we reset the cache into a random fresh state for the next round.
This process was repeated 1,000 times for each $|G|$.

In the following, we use a $G$ with a catching probability of 90\%. 
Depending on the target of the attack, lower or higher confidence levels may be chosen.
Afterward, we derive the complexity of constructing such an eviction set as the number of cache accesses needed for the construction.
All results can be found in a condensed form in \autoref{rprc:tab:ppp_eval_results} at the end of the section.

\noindent\textit{\textbf{1. \acf{RRP}:}}\\
Looking at \rrp in \autoref{rprc:fig:catch_prop_g}, we see a sharp increase in catching probability early on.
The $90\%$ mark is reached at $31$ addresses and continues rising until it reaches $100\%$ at $102$ addresses.
We now investigate the complexity of constructing such a generalized eviction set of size $31$ for \rrp caches. 
Therefore, we use the number of memory accesses \ppp needs to build the eviction set as a metric for the complexity.

\begin{figure}[ht!]
    \centering
    \includegraphics[width=\columnwidth]{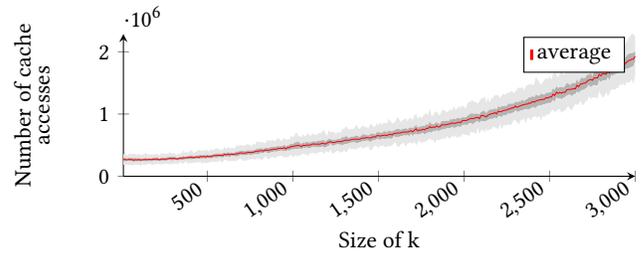}
    \caption{Required cache accesses to create an eviction set of size 31 for different pruning set sizes using \rrp.}
    \label{rprc:fig:randomrp_num_accesses}
\end{figure}

In \autoref{rprc:fig:randomrp_num_accesses}, the number of cache accesses for different sizes of the initial set $k$ measured over $1000$ iterations.
The dark grey area shows the interquartile range (middle 50\%), and the light gray area marks the boundary of the upper and lower whisker (i.e., the overall distribution with filtered outliers).

The results show that with \rrp, the best strategy for an attacker is using very small prime sets. 
The number of cache accesses is minimized at $|k|=110$ with 260,000 accesses.
The profiling succeeds even with very small $k$, i.e., $k=1$.
That is, since \rrp is stateless, and hence, it is possible that two consecutive cache accesses evict each other.
This would not be feasible using a \lru -like policy since the victim address would never be the immediate next replacement candidate.
Intuitively, the results hold since each accessed address has an equal probability of evicting $V$.
Thus, the attacker does not gain an advantage by accessing multiple addresses before triggering the access to $V$.
The reason why the cost is not minimized at $k=1$ is the additional cache access caused by the victim.

\noindent\textit{\textbf{2. \acf{LRU}:}}\\
We repeated the measurements described above for \lru, although it is not practical in real-world implementations due to its hardware implementation overhead.
Looking at \autoref{rprc:fig:catch_prop_g} again, the plot for \lru significantly differs from \rrp.
The catching probability for \lru overall is significantly lower.
Hence, a much larger eviction set $G$ is needed to catch accesses to $V$.
\lru reliably reaches the $90\%$ mark at $|G|=1010$.
This is to be expected since $G$ must occupy all candidate entries of $V$ for an access to be caught.
If there is a candidate entry of $V$ that is not occupied by addresses from $G$, it will always be prioritized using \lru since it is older than the addresses from $G$ that have just been primed and pruned by the attacker.

\begin{figure}[ht!]
    \centering
    \includegraphics[width=\columnwidth]{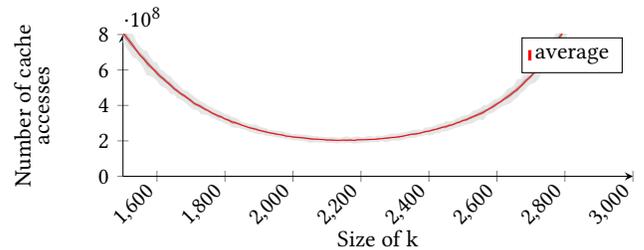}
    \caption{Required cache accesses to create an eviction set of size 1010 for different pruning set sizes using \lru.}
    \label{rprc:fig:lrurp_num_accesses}
\end{figure}

The complexity of creating an eviction set $G$ of size 1010 is illustrated in \autoref{rprc:fig:lrurp_num_accesses}.
Creating an eviction set with $|k| = 1$ was possible in an \rrp cache, but it is not feasible with \lru.
Again, that is, since all candidate entries of $V$ need to be filled by addresses from $k$. 
Hence, the minimum number of addresses required to catch a conflict is 4 in our evaluation environment.
However, this is very unlikely to occur.
In reality, much larger $k$'s are needed.
In our experiment, the smallest practical size of $k$ for \lru is approximately 1,500.
Decreasing the size will increase the number of missed accesses to $V$, leading to increased cache accesses needed to construct $G$.
The lowest average number of cache accesses in our experiment is 150 million for $|k| = 2,180$.
This is orders of magnitude more than for \rrp, indicating that it is substantially harder to find colliding addresses using \lru.

\noindent\textit{\textbf{3. \acf{DRPLRU}:}}\\
\autoref{rprc:fig:catch_prop_g} shows that the catching probability of \drplru behaves similarly to \rrp.
The $90\%$ mark is reached slightly quicker at $|G| = 29$.
Interestingly, the catch probability drops again for $|G| > 350$, implying that when using \drplru, it becomes harder to prune large sets of partial colliding addresses.

\begin{figure}[ht!]
    \centering
    \includegraphics[width=\columnwidth]{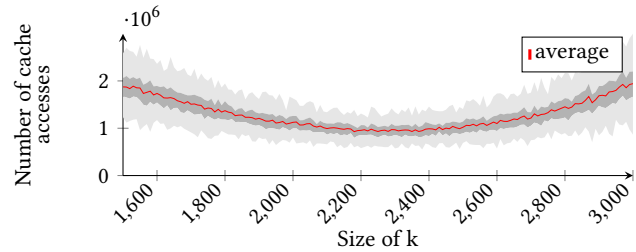}
    \caption{Required cache accesses to create an eviction set of size 29 for different pruning set sizes using \drplru.}
    \label{rprc:fig:rplrudynamicrp_num_accesses}
\end{figure}

The number of accesses required by the \ppp algorithm to construct an eviction set of size 29 is shown in \autoref{rprc:fig:rplrudynamicrp_num_accesses}.
Interestingly, in contrast to the catching probability, the required cache accesses \drplru develop more similar to \lru than \rrp. 
Similar to \lru, we found that for $|k|<1,500$, the required cache accesses increase steadily.
The minimum average number of cache accesses is reached at $|k| = 2,250$, with about 1 million accesses.
For larger $k$, the number of required cache accesses increases again.
Thus, constructing a reliable eviction set using \drplru requires three times more cache accesses than for \rrp but orders of magnitude fewer cache accesses than for the \lru replacement policy.

\noindent\textit{\textbf{4. \acf{FRPLRU}:}}\\
For \frplru, the 90\% threshold for a generalized eviction set is reached at $|G| = 570$, as shown in \autoref{rprc:fig:catch_prop_g}.
Therefore, the required eviction set size is much higher than for \rrp and \drplru but still by a factor of two smaller than for \lru.
Interestingly, the catching probability never exceeds 90\% by margin since, for larger $G$, the pruning fails more often.
Unlike with \lru replacement, catching an access to $V$ does not require occupying all possible entries for $V$.
It suffices to occupy one of those entries $e$ and access other addresses that map to the same set as $e$.
This effectively ages $e$, which makes it a possible candidate for replacement when $V$ is accessed.
This requires fewer addresses than occupying every possible entry for $V$.

\begin{figure}[ht!]
    \centering
    \includegraphics[width=\columnwidth]{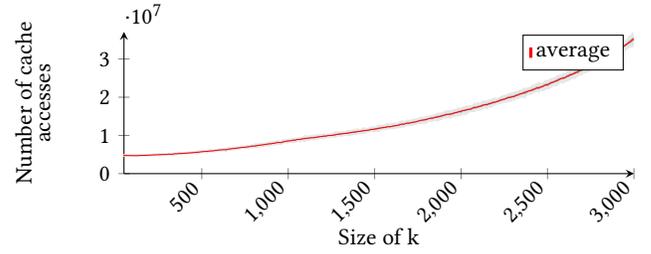}
    \caption{Required cache accesses to create an eviction set of size 570 for different pruning set sizes using \frplru.}
    \label{rprc:fig:rplrufixedrp_num_accesses}
\end{figure}

The number of cache accesses needed by \ppp to construct $G$ of size 570 with \frplru is shown in \autoref{rprc:fig:rplrufixedrp_num_accesses}.
Here, the characteristics are more similar to \rrp than \drplru and \lru, i.e., the profiling effort is minimized for small $k$.
As for \rrp, the number of cache accesses is minimized for $|k| = 110$.
At this point, we measured an average number of 4 million cache accesses which -- even though we are using a much smaller $k$ -- is by a factor of 4 higher than for \drplru.
The results indicate that accessing multiple addresses before triggering the access to $V$ does not significantly increase the chances of catching the access.
That is, after accessing one random address, it can become the next eviction candidate if, by chance, all other candidate entries have the lowest age in their cache set. 
In this situation, \frplru will randomly select one entry, which could be the currently used $k$ by chance.
In order to increase the chances of evicting an address of $k$, the attacker needs to access other entries at the same cache index.
Since the attacker has no better way of finding such addresses than accessing random addresses, a large $k$ is required for this to occur reliably.
As evident in \autoref{rprc:fig:rplrufixedrp_num_accesses}, the overhead introduced by such a large $k$ outweighs the benefit of a higher eviction probability.

Overall, \frplru requires significantly larger generalized eviction sets to catch an access \textit{and} has the highest measured profiling effort of the replacement policies suitable for randomized caches.
However, unlike \drplru, the attacker can succeed with very small $k$.
In noisy environments, pruning a large $k$ can be difficult.
Hence, \drplru may be preferable in such environments.

\noindent\textit{\textbf{5. \acf{VARP}:}}\\
We tested \varp with different age limits to see how this changes the probability of catching a victim access, the results of which can be found in \autoref{rprc:apx:varp_catch}.
In the following, we analyze \varp -64, i.e., a \varp implementation that allows 64 different age states per entry.
\autoref{rprc:fig:catch_prop_g} shows the catching probability for \varp with an age size of $64$.
The 90\% threshold is reached at 131 addresses in $G$.
The curve characteristics are similar to \rrp and \drplru though the catching probability rises much more slowly.
Opposed to \lru and \frplru, pruning large eviction sets does not cause many errors, and thus, a 100\% catch rate can be achieved.

\begin{figure}[ht]
    \centering
    \includegraphics[width=\columnwidth]{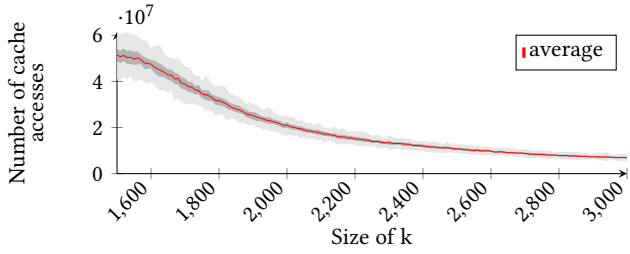}
    \caption{Required cache accesses to create an eviction set of size 131 for different pruning set sizes using \varp -64.}
    \label{rprc:fig:variableagerp_num_accesses}
\end{figure}

\autoref{rprc:fig:variableagerp_num_accesses} shows the average number of cache accesses needed to construct a generalized eviction set of size 131.
Opposed to the other replacement policies, \varp -64 favors large sizes of $k$ to minimize the overall accesses.
In our experiment, we reached the minimum number of cache accesses at $|k| = 3,000$ with 6 million accesses to construct $G$.
The reason that the attack on \varp works best for large $k$ is that any access will increase the age of all candidate entries.
Hence, by accessing many addresses, the attacker makes sure that many addresses of $k$ have a high age and are, therefore, likely to be replaced by the victim address.

\varp -64 requires the highest number of accesses to construct a reliable eviction set of all suitable candidates (excluding \lru).
Moreover, the large size of $k$ required to construct such an eviction set makes it difficult for an attacker to construct such an eviction set in a noisy environment.
An imaginary version of \varp with indefinitely large age indicators would perfectly resemble \lru since it would allow full access ordering.

\subsection{Targeted Evictions} \label{rprc:sec:targeted_evictions}
We now evaluate how the replacement policies affect the attacker's ability to evict entries from the cache.
As described in \autoref{rprc:sec:ppp} the \ppp attack~\cite{purnal2021systematic} requires the attacker to evict the victim address $V$ from the cache in each attack iteration. 
There are two approaches to accomplish this: Either access addresses randomly to evict $V$ by chance or access a generalized eviction set $G'$ of $V$ to evict it. 
Of course, combining both approaches is possible, i.e., accessing several colliding addresses followed by several random addresses to evict $V$ or vice versa.
Since the attacker has no way of learning when the eviction of $V$ is successful -- that would require probing $V$, which brings it back into the cache -- they need to choose the number of colliding- and random addresses such that the probability of evicting $V$ is substantial.
Otherwise, the following iteration of \ppp is guaranteed to fail.
In the following, we use a targeted success probability of 50\%.
We found that evicting $V$ from the cache requires significantly more cache accesses than catching an access to $V$.

In this section, we compare the number of required accesses to random- and colliding addresses for the eviction of $V$ for each replacement policy.
We then derive the complexity of constructing a sufficiently sized generalized eviction set of $V$ and discuss whether the benefit of colliding addresses outweighs the overhead of constructing a generalized eviction set.

\noindent\textit{\textbf{1. \acf{RRP}:}}
\begin{figure}[ht!]
    \centering
    \includegraphics[width=.8\columnwidth]{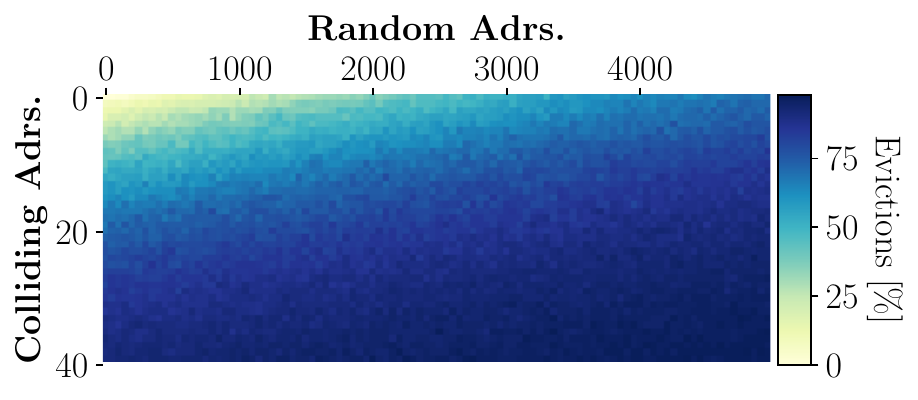}
    \caption{Number of cache accesses with random- or colliding addresses and combinations (random $\rightarrow$ colliding) thereof until the victim entry is replaced using \textit{random replacement policy}.}
    \label{rprc:fig:rp_single_evict_rand}
\end{figure}

\noindent \autoref{rprc:fig:rp_single_evict_rand} shows the eviction characteristics of \rrp.
For this, we performed 500 iterations of accessing $y$ random addresses followed by $x$ colliding addresses for each $x$ and $y$.
We then plot the probability that these accesses evict a randomly selected $V$.
Note that for some replacement policies, the results depend on the order of accessing random and colliding addresses.
We generally found that starting with random addresses and then using colliding addresses increases the attackers' success probability; see details in \autoref{rprc:apx:access}.
The figure shows that the attacker needs significantly fewer colliding addresses than random addresses to evict $V$.
For example, using 11 colliding addresses results in an eviction probability of 50\%, the same as using 2,900 random addresses.
This is expected since \rrp is stateless and each accessed address $a$ has a 1-in-$w$ chance of evicting $V$ if $a$ collides with the entry of $V$.
In our experiment, the \ppp algorithm, on average, requires 92,284 cache accesses to build such an eviction set.
Note that $G$ and $G'$ must be disjunctive since, after the attack, $G$ will be cached and, therefore, cannot evict $V$.

Comparing the number of addresses needed to create $G'$ of size 11 with the number of accesses to random addresses required to achieve the same, we conclude that when \rrp is employed in the cache, in most cases, it is beneficial to invest in creating $G'$ than having to access 2,900 random addresses in every iteration.

\noindent\textit{\textbf{2. \acf{LRU}:}}
\begin{figure}[ht!]
    \centering
    \includegraphics[width=.8\columnwidth]{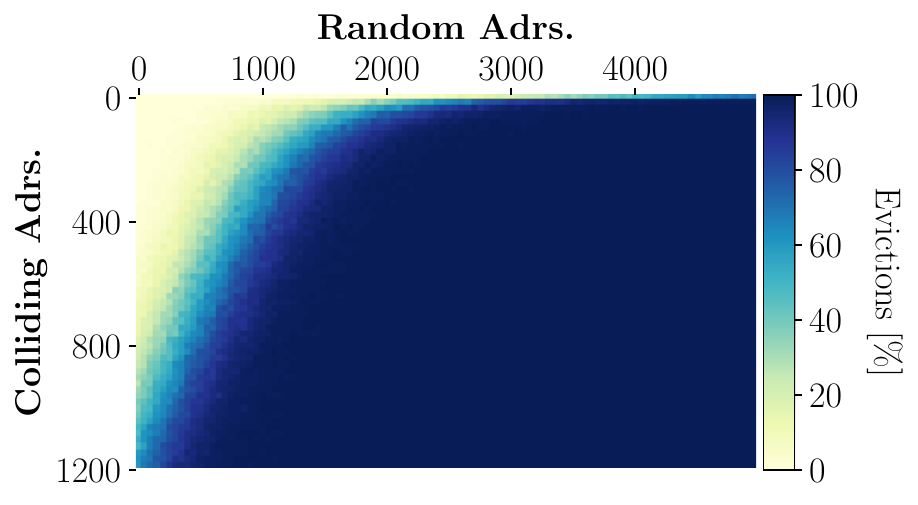}
    \caption{Number of cache accesses with random- or colliding addresses and combinations (random $\rightarrow$ colliding) thereof until the victim entry is replaced using \textit{\lru}. Note the different y-axis scaling compared to Figures \ref{rprc:fig:rp_single_evict_rand} and \ref{rprc:fig:rp_single_evict_rplrudyn}.}
    \label{rprc:fig:rp_single_evict_lru}
\end{figure}

\noindent Though \lru is not practical for randomized caches in real-world implementations, we repeated the above experiment with an ideal \lru cache.
The results are shown in \autoref{rprc:fig:rp_single_evict_lru} and drastically differ from the results of \rrp.
In particular, the attacker requires significantly more addresses to evict $V$ from the cache.
When accessing random addresses, the attacker must access more than 5,000 addresses to have a decent probability of evicting $V$.
That is because for the victim to be replaced, the accessed address $a$ not only needs to collide with $V$ in at least one way, but also all other candidate entries of $a$ need to have been accessed after the victim. 
Since the collection of candidates is (pseudo-) random, the attacker needs to occupy a large part of the cache for this to occur.
Accessing a set of random addresses followed by colliding addresses improves the attackers' success probability immensely. 
However, the amount of colliding addresses required is huge, as analyzed in the following.

To achieve a 50\% eviction probability, the attacker would need to access as many as 1,010 colliding addresses.
However, \autoref{rprc:fig:rp_single_evict_lru} shows that the size of the eviction set can be reduced by accessing random addresses before accessing the set.
For example, when accessing 1,000 random addresses, the attacker would only need an eviction set $G'$ of size 400 to reach the same eviction probability.
Therefore, for \lru caches, mixing random- and colliding addresses is advantageous compared to the extensive profiling required to generate $G'$.

\noindent\textit{\textbf{3. \acf{DRPLRU}:}}
\begin{figure}[ht!]
    \centering
    \includegraphics[width=.8\columnwidth]{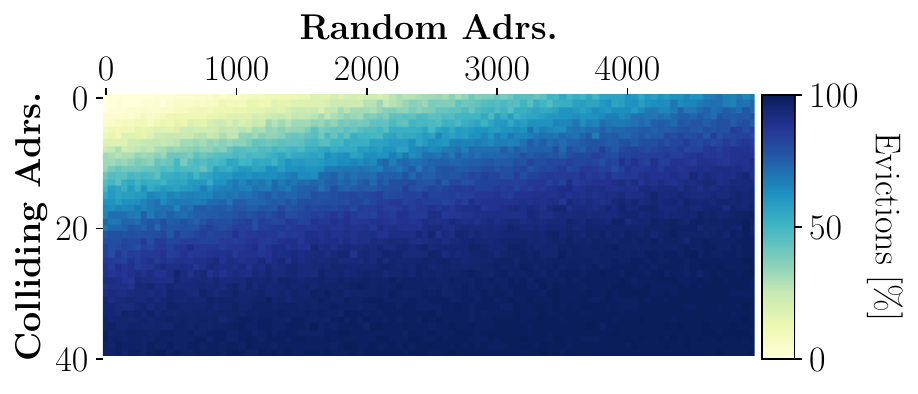}
    \caption{Number of cache accesses with random- or colliding addresses and combinations (random $\rightarrow$ colliding) thereof until the victim entry is replaced using \textit{\drplru}.}
    \label{rprc:fig:rp_single_evict_rplrudyn}
\end{figure}

\noindent The eviction results for \drplru are shown in \autoref{rprc:fig:rp_single_evict_rplrudyn}.
The relation of random addresses and colliding addresses is similar to the results using \rrp as replacement policy.
However, the attacker must access more addresses to evict $V$ than with \rrp. 
The attacker has a 50\% success rate after accessing 16 colliding addresses (compared to 11 for \rrp) or 3,400 random addresses (compared to 2,900 for \rrp).
Thus, in this regard, \drplru slightly outperforms \rrp though it remains in the same order of magnitude.
However, an important observation comparing \rrp and \drplru is that \drplru is less likely to evict $V$ after very few accesses but, therefore, more likely to evict $V$ after many accesses.
This means that the transition from low to high probability of success is sharper for \drplru than for \rrp.
Hence, if the attacker targets a high eviction rate, \drplru is actually less secure than \rrp.

On average, \ppp needs to access 513,681 addresses to build a $G'$ of size 16 for \drplru.
Compared to \rrp, where the sizes of the eviction sets are similar (11 for \rrp), the profiling complexity is five times as high.
Since accesses to random addresses do not influence the eviction probability as much as in \lru and the required size of $G'$ is small, constructing $G'$ prior to the attack can be beneficial.

\noindent\textit{\textbf{4. \acf{FRPLRU}:}}
\begin{figure}[ht!]
    \centering
    \includegraphics[width=.8\columnwidth]{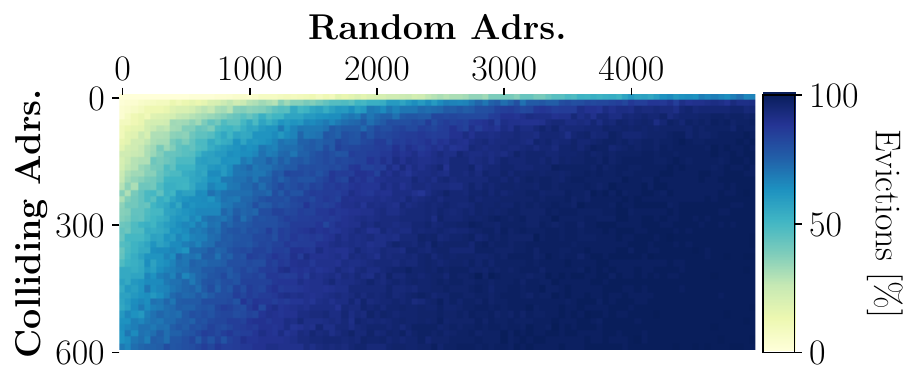}
    \caption{Number of cache accesses with random- or colliding addresses and combinations (random $\rightarrow$ colliding) thereof until the victim entry is replaced using \textit{\frplru}. Note the different y-axis scaling compared to Figures~\ref{rprc:fig:rp_single_evict_rand} and~\ref{rprc:fig:rp_single_evict_rplrudyn}.}
    \label{rprc:fig:rp_single_evict_rplrufix}
\end{figure}

\noindent 
The results of \frplru vary significantly from those of \drplru, as shown in \autoref{rprc:fig:rp_single_evict_rplrufix}.
Like \lru, \frplru requires a substantially higher number of accesses to evict $V$ than \rrp and \drplru.
However, compared to \lru, the number of required addresses is still lower.
Using random addresses only, the attacker needs more than 5,000 accesses to reliably evict $v$.
Accessing colliding addresses accelerates the eviction of $V$, though the attacker is still required to access 520 colliding addresses for a 50\% success rate of evicting $V$.
This is because, for a high eviction probability of $V$, the attacker first needs to increase the age of $V$. 
Since the age is maintained relative to other entries that share the same cache index, the attacker thus needs to replace those entries first.
For this, colliding addresses have no advantage over random addresses.
Hence, by accessing random addresses, the attacker brings the victim entry $V$ into a state where it will likely be replaced when accessing colliding addresses.

Using \ppp to profile the cache and create $G'$ of size 520 would, on average, require 4.3 million addresses to be accessed in our evaluation environment.
A combination of random- and colliding addresses to evict $V$ reduces the required size of $G'$.
For example, using 1,000 random addresses during the eviction process reduces the required size of $G'$ to 360 for a 50\% success rate in our environment.
Regardless of the strategy, the attacker needs to access a vast number of addresses in each iteration, reducing the attack granularity due to long probing intervals.

\noindent\textit{\textbf{5. \acf{VARP}:}}
\begin{figure}[ht!]
    \centering
    \includegraphics[width=.8\columnwidth]{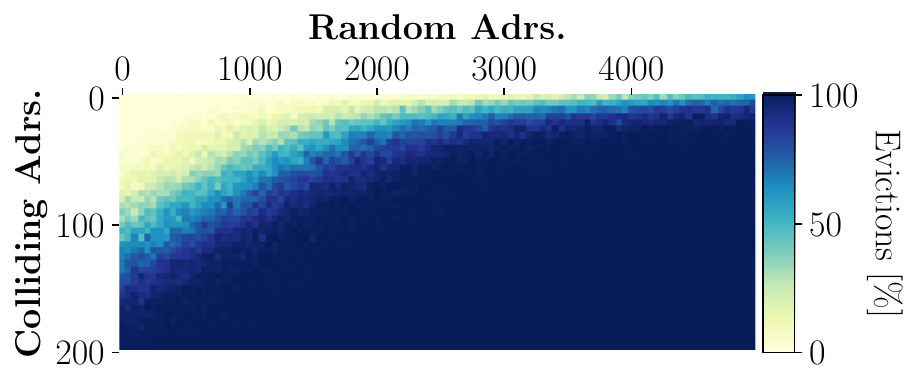}
    \caption{Number of cache accesses with random- or colliding addresses and combinations (random $\rightarrow$ colliding) thereof until the victim entry is replaced using \textit{\varp -64}.}
    \label{rprc:fig:rp_single_evict_varage}
\end{figure}

\noindent
In contrast to \frplru and \rrp, the eviction results for the \varp -64 replacement policy show a sharp transition from low- to high success probability.
The attacker achieves a 50\% eviction rate by accessing 125 colliding addresses while needing over 5,000 random addresses to achieve similar results.
This is the third highest value for the colliding addresses after \lru and \frplru.
We also see that random addresses do not influence the eviction probability as much as in \lru and \frplru.
This means the attacker cannot effectively trade more random accesses for a reduced eviction set.

Using the values from the previous section, we can calculate that \varp -64 ages require, on average, 6.5 million cache accesses to create $G'$ with size 125.
This value is higher than for \frplru, even though only $25\%$ of the colliding addresses are required.
As a result, we infer that it is more challenging to build an eviction set using \varp -64 than \frplru.

\begin{table}
\centering
    \addtolength{\leftskip} {-2cm}
    \addtolength{\rightskip}{-2cm}
\footnotesize
    \caption{Summary of the \ppp attack evaluation using different replacement policies.}
    \label{rprc:tab:ppp_eval_results}
    \begin{tabular}{p{2cm}rrrrr}
        \toprule
        Policy                     & Random  & LRU         & FRPLRU    & DRPLRU  & VARP-64   \\\midrule
        Optimal $|k|$                          & 110     & 2180        & 110       & 2250    & 3000      \\
        $|G|$ for $90\%$ catch probability     & 31      & 1010        & 570       & 29      & 131       \\        
        Accesses to create $G$           & 260,074 & 150,430,216 & 4,714,267 & 931,047 & 6,840,702 \\
        $|G'|$ for $50\%$ eviction probability & 11      & 1010        & 520       & 16      & 125       \\
        Accesses to create $G'$          & 92,284  & 150,430,216 & 4,300,734 & 513,681 & 6,527,387 \\
        \bottomrule
    \end{tabular}
\end{table}

\subsection{Comparison and Summary}
We summarize our findings from the previous two sections in \autoref{rprc:tab:ppp_eval_results}.
From the evaluation of the catching probability, we learned that \lru needs the most colliding addresses followed by \frplru, \varp -64, \rrp, and \drplru to catch accesses to a victim address reliably.
Moreover, we found that for most tested replacement policies, the eviction of $V$ after each iteration is more challenging than actually catching an access to $V$.
That is, eviction set sizes allowing a 90\% catch rate are only sufficient to evict $V$ with a $50\%$ success probability for \lru, \frplru, and \varp -64. 
In contrast, only half the addresses are needed for \rrp and \drplru.
Note that $G$ and $G'$ need to be disjunctive since after the attack, addresses from $G$ will be cached and, therefore, not be useful to evict $V$.
Examining the heatmaps in the previous section, we see that the transitions from low to high eviction probabilities are unevenly sharp for the replacement policies.
For \rrp and \drplru, the transitions are smooth, meaning if the attacker needs an eviction probability of around $90\%$, the required number of addresses will rise significantly.
For \lru and \varp -64, these transitions are much sharper.
Trying to reach a $90\%$ eviction probability would require much fewer additional addresses than for \rrp and \drplru.
The exception from this is \frplru.
We see that the required sizes for $G$ with a 90\% catch rate and $G'$ with a 50\% eviction rate are almost equal, but the transition from low to high eviction probabilities is smooth.
This indicates that \frplru requires significantly more addresses for a high eviction rate.

Looking at the complexity of creating such $G$ and $G'$, we see that \lru has the highest complexity by far, with over 150 million cache accesses, but it also needs the largest $G$.
In contrast, \rrp requires only a fraction of the accesses.
While \rrp and \drplru require a similar number of colliding addresses, \drplru still has a higher complexity.
This results from the 20-times larger $k$ for \drplru, leading to more cache accesses overall.
The same goes for \frplru and \varp, which both need larger $G$ and $G'$ and have a higher complexity indicating a higher level of security.
\frplru requires more than four times the colliding addresses while featuring only 30$\%$ of the complexity of \varp -64.
Again, this is the result of \frplru requiring a 27-times smaller $k$ used in \ppp.

\section{Performance Evaluation}

We use the gem5 simulator \cite{lowepower2020gem5}, introduced in \autoref{rprc:sec:eval_env}, to implement and simulate the replacement policies.
The performance is measured by running the PARSEC benchmark \cite{bienia11benchmarking} in gem5's full-system mode.
gem5 is configured to use a two-level cache hierarchy consisting of an L1 and an L2 cache.
The replacement policies only apply to the L2 cache, while the L1 data and instruction caches use \lru replacement.
Focusing only on data caches, the L1 is 64 KiB large with 256 sets and four ways of associativity.
The L2 cache has a total size of 256 KiB with 1024 sets and four ways of associativity.

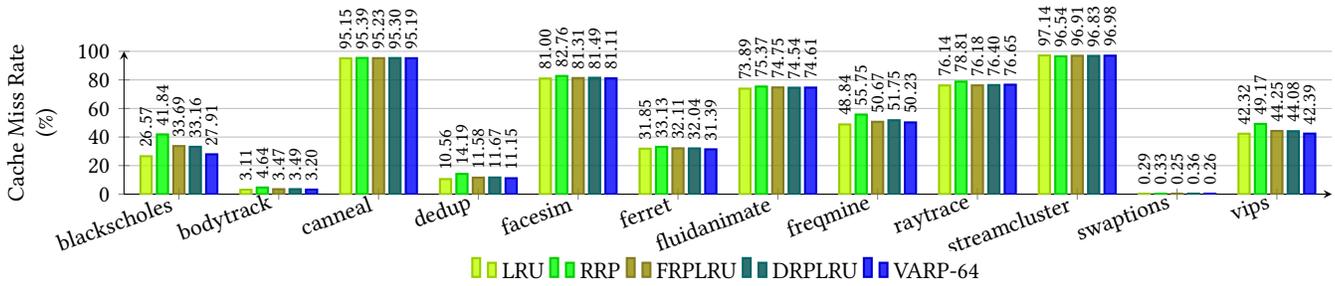
\begin{figure*}[t!]
    \centering
    \definecolor{black_coral}{HTML}{5b616a}
\begin{tikzpicture}
    \begin{axis}[
            ybar,
            width = \textwidth,
            height = 3.5cm,
            axis lines=left,
            bar width=0.15cm,
            enlarge x limits=0.05,
            xtick=data,
            xticklabels={blackscholes, bodytrack, canneal, dedup, facesim, ferret, fluidanimate, freqmine, raytrace, streamcluster, swaptions, vips, x264},
            x tick label style={rotate=20,anchor=east},
            ymajorgrids=true,
            ymin=0,
            nodes near coords,
            nodes near coords align={horizontal},
            every node near coord/.append style={font=\footnotesize, color=black, opacity=1, rotate=90},
            nodes near coords={\pgfmathprintnumber[fixed,fixed zerofill,precision=2]{\pgfplotspointmeta}},
            ymax = 100,
            ylabel style={align=center},
            ylabel = {Cache Miss Rate\\ (\%)},
            legend style={
                    draw = none,
                    at={(0.5,-0.4)},
                    anchor=north,legend columns=5},
        ]

        \addplot[lime!80!black, thick,fill=lime,fill opacity=.8] coordinates {(0, 26.5671) (1, 3.1102000000000003) (2, 95.1478) (3, 10.5585) (4, 81.0028) (5, 31.8454) (6, 73.8866) (7, 48.8375) (8, 76.1379) (9, 97.1392) (10, 0.2932) (11, 42.322700000000005) };
        \addlegendentry{\lru}
        \addplot[green!80!black, thick,fill=green,fill opacity=.8] coordinates {(0, 41.8358) (1, 4.643) (2, 95.3928) (3, 14.190900000000001) (4, 82.7604) (5, 33.1347) (6, 75.366) (7, 55.745599999999996) (8, 78.8104) (9, 96.544) (10, 0.3274) (11, 49.172) };
        \addlegendentry{\rrp}
        \addplot[olive!80!black, thick,fill=olive,fill opacity=.8] coordinates {(0, 33.6877) (1, 3.4703) (2, 95.2317) (3, 11.5769) (4, 81.31479999999999) (5, 32.1056) (6, 74.7542) (7, 50.6699) (8, 76.1797) (9, 96.9062) (10, 0.2523) (11, 44.2543) };
        \addlegendentry{\frplru}
        \addplot[teal!80!black, thick,fill=teal!60!black,fill opacity=.8] coordinates {(0, 33.1633) (1, 3.4948) (2, 95.3005) (3, 11.665000000000001) (4, 81.48769999999999) (5, 32.0389) (6, 74.543) (7, 51.7528) (8, 76.3998) (9, 96.833) (10, 0.363) (11, 44.083800000000004) };
        \addlegendentry{\drplru}
        \addplot[blue!80!black, thick,fill=blue,fill opacity=.8] coordinates {(0, 27.9127) (1, 3.2006) (2, 95.1922) (3, 11.1491) (4, 81.109) (5, 31.3898) (6, 74.6142) (7, 50.2287) (8, 76.6467) (9, 96.9774) (10, 0.2611) (11, 42.3942) };
        \addlegendentry{\varp -64}
    \end{axis}
\end{tikzpicture}
    \caption{\label{rprc:fig:rpparsec}Cache miss rate statistics of the PARSEC benchmark suite.}
\end{figure*}

We run each PARSEC workload for each replacement policy with the predefined medium-sized workload input \textit{simmedium}.
We use the cache miss rate of the L2 cache as a metric for our evaluation.
\autoref{rprc:fig:rpparsec} shows the cache miss rate for each workload and replacement policy combination.
The results are grouped by workloads as they have different memory access behaviors, and thus the results are only comparable for one workload.
A higher cache miss rate indicates worse performance.

\rrp has the highest cache miss rate for every workload except \textit{streamcluster} and \textit{swaptions}.
This was expected since replacing cache entries randomly might replace frequently used ones leading to increased cache misses.
\lru and \varp performed best, with \lru being slightly ahead of \varp.
The overall performance difference between the policies heavily depends on the used workload.
For example, for \textit{canneal}, the difference is bearly noticeable, with only $0.24\%$.
In contrast, for \textit{blackscholes}, the difference between the best and worst performing policy is $15.27\%$.
\drplru and \frplru perform equally over all workloads.
The only exception is \textit{freqmine}, where the miss rate for \frplru is around $1\%$ lower.

\section{Area Evaluation}

This section evaluates the proposed replacement policies in terms of their overall overhead.
We implemented our replacement policies on a Xilinx Kintex-7 FPGA chip using Vivado 2022.2.
The post-synthesis area results are shown in Table \ref{tab:synthArea}.

\begin{table}[htbp]
\small
    \centering
    \caption{Post-synthesis area results on Kintex-7 FPGA}
    \label{tab:synthArea}
    \begin{tabular}{lccccc}
     \toprule
     & BRAMs & LUTs & \% of cache & FFs & \% of cache \\\midrule
    \small{CV32E40P}       & -  & 4950 &      & 2142 &      \\ 
    \small{Core+cache}     & 18 & 7678 &      & 3105 &      \\ \midrule
    \small{RRP}            & -  & 26   & 0.95 & 16   & 1.66 \\
    \small{DRPLRU}         & 2  & 55   & 2.02 & 27   & 2.80 \\
    \small{FRPLRU}         & 2  & 26   & 0.95 & 18   & 1.87 \\
    \small{VARP-64}        & 2  & 80   & 2.93 & 43   & 4.46 \\\bottomrule
    \end{tabular}
\end{table}

Table \ref{tab:synthArea} shows the core and CPU (CV32E40P core + cache) areas in rows 1 and 2, respectively.
We use these results in order to compare the cost of replacement policies with the cache and the system.
We include SCARF modules in the cache area.
SCARF modules make up a large proportion of the cache area, representing ~80\% of the LUTs and ~6\% of the FF registers.
Then, for rows 3 to 6, Table \ref{tab:synthArea} shows the cost of the replacement policies in the cache.
For the RRP, we retrieve the 16 FF registers that are used to store the 16-bit depth LFSR.
We notice that \drplru, \frplru, and \varp -64 use 2 Block RAMs to store the state values of each cache line.
All replacement policies use few hardware resources (from 0.95 to 2.93\% of cache LUTs, and 1.66 to 4.46\% of cache FF registers).
However, we can extract some trends in terms of the area alone.
RRP uses the least resources in terms of LUTs and registers and does not require Block RAM. 
The three lasts require BRAMs, of which VARP-64 uses the most resources over DRPLRU and FRPLRU.
To conclude, the hardware resources used by the replacement policies are negligible.
Especially as we assess it with a basic cache created for a small processor, the cost of the hardware will further decrease upon reviewing it in the context of a larger system.
\section{Related Work}
\lru and \rrp are two well-known replacement policies.
As \lru is expensive to implement for high associativities, the focus shifted to Pseudo-LRU policies approximating \lru.
Approaches based on binary trees can be implemented with lower overhead but do not approximate \lru very well, as shown by Robinson \cite{Robinson2004}.
He found that using non-binary trees can lead to a better approximation of \lru.
Another approach was by Jaleel \etal \cite{Jaleel2010} in 2010.
They proposed \ac{RRIP}, which uses $M$ bits to store Re-Reference Prediction Values per cache entry.
Those values indicate if an entry will be re-referenced in the near-immediate (0) or distant ($2^M-1$) future.
On a cache-miss, those entries with a predicted re-reference in the distant future are replaced.
In \cite{Wu2011,Young2017}, replacement policies based on the PC are proposed.
They use the PC-based signatures to predict re-reference behavior of cache entries.
In 2021 Sethumurugan \etal \cite{Sethumurugan2021} showed that it is possible to apply machine learning techniques to replacement policy design. 

Cache randomization has been introduced as a measure against side-channel attacks by Wang and Lee~\cite{DBLP:conf/isca/WangL07,Wang2008}.
CEASER~\cite{DBLP:conf/micro/Qureshi18} and CEASER-S~\cite{DBLP:conf/isca/Qureshi19} use an encryption mechanism for efficient randomization.
Suitable randomization schemes have been constructed and analyzed in~\cite{Canale2023,Bodduna2020}.
ScatterCache~\cite{Werner2019} and PhantomCache~\cite{DBLP:conf/ndss/TanZB020} provide an increased randomization space, thus increasing the attack resistance.
Purnal \etal~\cite{purnal2021systematic} analyze the security of randomized cache architecture and propose the \ppp attack, which targets generic randomized cache architectures.
Mirage~\cite{DBLP:conf/uss/SaileshwarQ21} and ClepsydraCache~\cite{Thoma2023} implement additional measures to prevent such attacks.

\section{Conclusion}
In this paper, we analyzed suitable replacement policies for randomized cache architectures.
We choose the two well-known replacement policies, \lru and \rrp, the in the context of \ac{TLB}s proposed \frplru replacement policy and two new replacement policies, \drplru and \varp.
These were evaluated regarding their impact on a state-of-the-art cache attack \ppp and their impact on the overall cache performance.
Additionally, we implemented a randomized cache using SCARF \cite{Canale2023} and the replacement policies in hardware on a CV32E40P RISC-V core.
This allowed us to retrieve area results for the randomized cache and the replacement policies.
While, according to our results, \lru is the most secure replacement policy, it is also impractical to implement in hardware.
However, our results also show that alternatives exist (\frplru, \varp -64) that feature a sufficiently high level of security while keeping the performance and area overhead minimal.

\begin{acks}
The work described in this paper has been supported by the Deutsche Forschungsgemeinschaft (DFG, German Research Foundation) under Germany's Excellence Strategy - EXC 2092 CASA - 390781972 and under the Priority Program SPP 2253 Nano Security (Project RAINCOAT - Number: 440059533).
It was also supported by the Cominlabs excellence laboratory with funding from the French National Research Agency (ANR-10-LABX-07- 01), the Collège Doctoral de Bretagne, and the research group GDR ISIS. 
\end{acks}

\bibliographystyle{ACM-Reference-Format}
\bibliography{main.bib}

\appendix

\section{Access Order}
\label{rprc:apx:access}

In this Appendix, we analyze the impact of the access order of random- and colliding addresses on the eviction rate of $V$ for different replacement policies (\ding{183}).

\begin{figure}[ht!]
    \centering
    \includegraphics[width=\columnwidth]{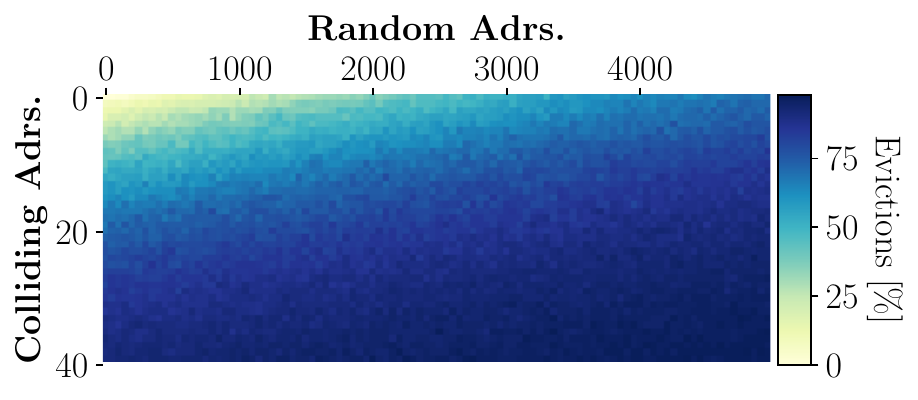}
    \caption{Number of cache accesses with random- or colliding addresses and combinations (random $\rightarrow$ colliding) thereof until the victim entry is replaced using \textit{\rrp}.}
    \label{rprc:fig:rp_single_evict_RRP_APP}
\end{figure}

\begin{figure}[ht!]
    \centering
    \includegraphics[width=\columnwidth]{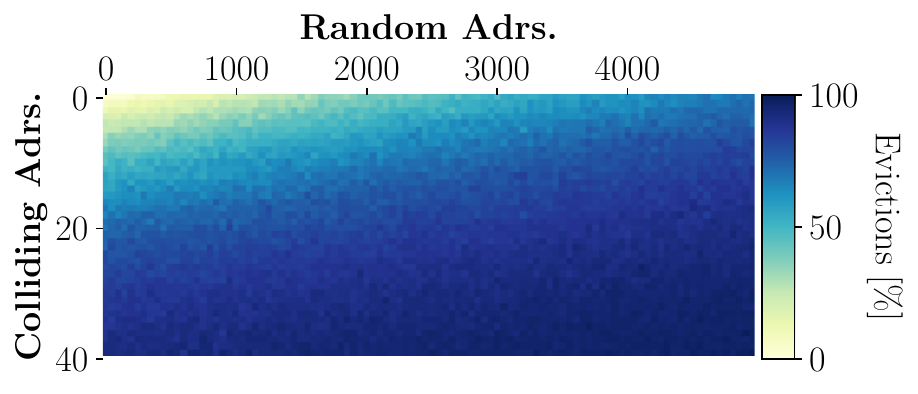}
    \caption{Number of cache accesses with random- or colliding addresses and combinations (colliding $\rightarrow$ random) thereof until the victim entry is replaced using \textit{\rrp}.}
    \label{rprc:fig:rp_single_evict_RRP2_APP}
\end{figure}

For \rrp and \drplru, we found that there is no difference in the eviction probability regardless of the order of accessing random- and colliding addresses. 
\autoref{rprc:fig:rp_single_evict_RRP_APP} shows the eviction probability of an \rrp cache using colliding addresses followed by random addresses, and \autoref{rprc:fig:rp_single_evict_RRP2_APP} shows the inverted case.
There is no significant difference in the results which indicates that the access order does not matter.
The results for \drplru are similar.

\begin{figure}[ht!]
    \centering
    \includegraphics[width=\columnwidth]{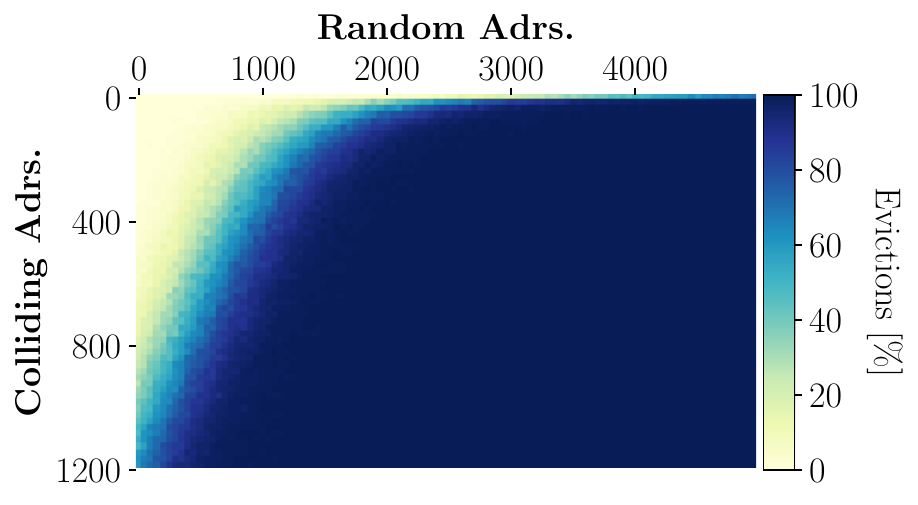}
    \caption{Number of cache accesses with random- or colliding addresses and combinations (random $\rightarrow$ colliding) thereof until the victim entry is replaced using \textit{\lru}.}
    \label{rprc:fig:rp_single_evict_LRU_APP}
\end{figure}

\begin{figure}[ht!]
    \centering
    \includegraphics[width=\columnwidth]{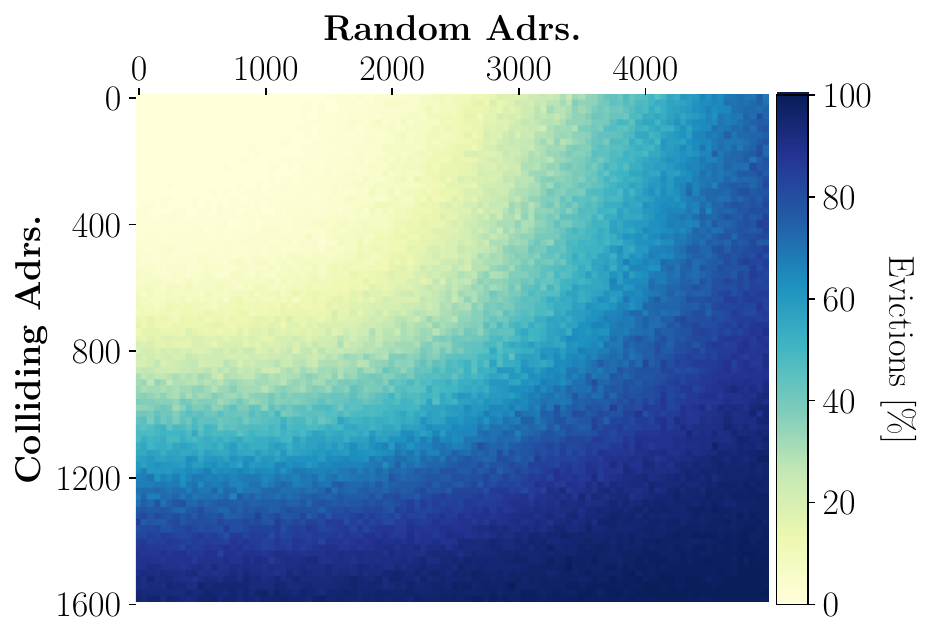}
    \caption{Number of cache accesses with random- or colliding addresses and combinations (colliding $\rightarrow$ random) thereof until the victim entry is replaced using \textit{\lru}.}
    \label{rprc:fig:rp_single_evict_LRU2_APP}
\end{figure}

For \lru, \frplru, and \varp -64, we found that the access order of random- and colliding addresses does matter.
For example, \autoref{rprc:fig:rp_single_evict_LRU_APP} shows the eviction probability for accessing random addresses followed by colliding ones, while \autoref{rprc:fig:rp_single_evict_LRU2_APP} shows the inverse access order case.
The difference is very clear, and the results show that the eviction probability is clearly better when first accessing random addresses followed by colliding ones. 
The reason for this is that when the attacker starts the eviction using colliding addresses, these will never evict $V$ since $V$ is still in a relatively new replacement state.
Thus, the actually known-to-colliding addresses are wasted since they are almost guaranteed not to evict $V$. 
By first accessing random addresses, the attacker increases the relative age of $V$ compared to other entries, which makes it more likely to become the eviction candidate when it is eventually accessed using know-to-collide addresses.

%
%
\section{Catching Probability for Different \varp Instances}
\label{rprc:apx:varp_catch}

\begin{figure}[ht!]
    \centering
    \includegraphics[width=\columnwidth]{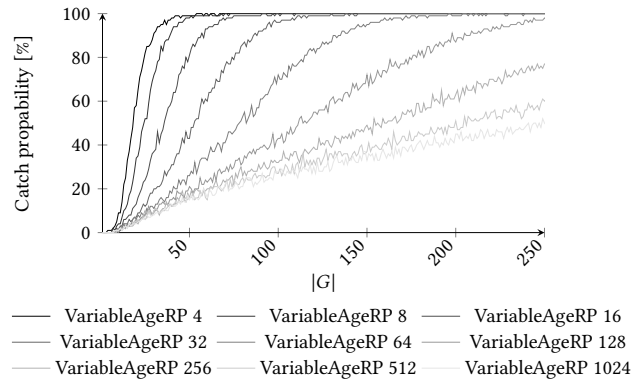}
    \caption{Catching probability for \varp with different age sizes as a function over the size of the eviction set $G$.}
    \label{rprc:fig:catch_prob_varp_multi}
\end{figure}

The primary aim of \varp is to investigate the behavior of a replacement policy using mostly independent age identifiers with variable sizes.
\autoref{rprc:fig:catch_prob_varp_multi} shows the development of the catching probabilities for different instances of \varp.
When using a small number of distinct age identifiers, the catching probability develops nearly identically with \rrp and \drplru.
This happens because the likelihood of two or more cache entries sharing the same age is high when selected by the indexing function.
Remember that \varp chooses one entry at random if there exist multiple entries sharing the same age.
As a result, \varp tends to make a high number of random choices.
However, as the number of age identifiers increases, the number of entries sharing the same age will decrease.
This decrease is accompanied by a reduction in the number of random decisions made by \varp.
By reducing the number of random choices and enforcing more statefulness, reflected in the figure as decreasing catch probability, \varp's catching probability profile becomes closer to LRU.

\end{document}